\documentclass[%
 reprint,
 amsmath,amssymb,
 aps,
pra,
floatfix,
]{revtex4-1}
\usepackage{graphicx,lipsum}
\usepackage{dcolumn}
\usepackage{bm}
\usepackage[dvipsnames]{xcolor}
\usepackage{array,multirow}
\usepackage{bm}
\usepackage{bbm}

\begin{document}

\preprint{APS/123-QED}
\title{The effect of quantum memory on quantum speed limit time for CP-(in)divisible channels}

\author{K.G. Paulson\textsuperscript{a}}
\altaffiliation[paulsonkgeorg@gmail.com]{}
\author{Subhashish Banerjee\textsuperscript{b}}
\email{subhashish@iitj.ac.in }
\author{R. Srikanth\textsuperscript{c}}
 \email{srik@poornaprajna.org }
\affiliation{Institute of Physics, Bhubaneswar-751005, India\textsuperscript{a}\\
Indian Institute of Technology Jodhpur, Jodhpur-342011, India\textsuperscript{b}\\
Poornaprajna Institute of Scientific Research, Sadashivnagar, Bengaluru-560080, India\textsuperscript{c}}



\date{\today}

\begin{abstract}
Quantum speed limit time defines the limit on the minimum time required for a quantum system to evolve between two states.  Investigation of bounds on speed limit time of quantum system under non-unitary evolution is of fundamental interest, as it reveals  interesting connections to quantum (non-)Markovianity. Here, we discuss the characteristics of quantum speed limit time as a function of quantum memory, quantified as the deviation from temporal self-similarity of quantum dynamical maps for CP-divisible as well as indivisible maps, and show that the presence of quantum memory can speed up quantum evolution. This demonstrates the enhancement of the speed of quantum evolution in the presence of quantum memory for a wider class of channels than indicated by the CP-indivisibility criterion.

\end{abstract}

\maketitle


\section{\label{sec:level1} Introduction }  
The dynamical characteristics of open quantum systems (OQS), a quantum system coupled to its environment, have received wide attention \cite{breuer2002,sbbook}. The coupling of a quantum system to its environment gives rise to realistically unavoidable processes, such as dissipation and decoherence. Memory effects lead to the evolution of the system of interest being classified as  (non-)Markovian. Due to theoretical as well as technological advances, the study of non-Markovian phenomena has attracted much attention in recent times \cite{kumar2018en,thomas2018,Shrikant2018,kumar2018,bhattacharya2018,naikoo2019,utagi2020ping,naikoo2020}. Typically, in a non-Markovian evolution,  the system and environment time scales are not well separated, which could lead to information backflow from the environment to the system \cite{rivas2014,breuer2016,de2017,li2018}. It is known that a quantum system coupled to a non-Markovian environment can evolve faster than the corresponding case of the Markovian environment~\cite{deffner2013quantum,liu2016}. This has been experimentally verified in a cavity  QED system~\cite{cimmarusti2015}. Since a study of quantum speed limit is helpful in obtaining physical insight into the dynamical properties of quantum systems, it would be worthwhile to make a detailed understanding of the quantum speed limit with memory.\newline
Heisenberg's energy-time uncertainty principle sets the bound on the minimum time required to evolve between two quantum states. Mandelstam and Tamm (MT)~\cite{mandelstam1945} provided an interpretation of energy-time uncertainty with the time scale of the quantum state's evolution. Later, Margolus and Levitin (ML)~\cite{margolus1998} derived an alternate bound for the minimal time evolution of a quantum system between two orthogonal states. The combined bound of $\tau_{QSL}$ quantum speed limit time, which is tight~\cite{levitin2009} for a closed quantum system, is  $\max\Bigg\{\frac{\pi\hbar}{2\Delta H},\frac{\pi\hbar}{2\langle H\rangle-E_0}\Bigg\}$. Here, $\Delta H$ and $\langle H\rangle$ are the square roots of the variance, and the expectation value of the Hamiltonian with respect to the initial state, respectively. The zero of the energy is generally considered as the ground state such that $E_0=0$. The bound on speed limit time was further refined for initial mixed, and driven systems for arbitrary angles of separation ~\cite{deffner2017,giovannetti2003}. In~\cite{uhlmann1992,deffner2013}  MT and ML bounds were generalized to arbitrary initial and final  mixed  states based on geometrical distance between them for time independent Hamiltonian , and is $\max\Bigg\{\frac{\hbar}{\Delta H}\mathcal{B}(\rho_{0},\rho_{\tau}),\frac{2\hbar}{\pi\langle H\rangle}\mathcal{B}(\rho_{0},\rho_{\tau})^2 \Bigg\}$, where  $\mathcal{B}(\rho_{0},\rho_{\tau})$ is the Bures angle between the initial and final states.

In reality, quantum systems are not isolated; interactions of the system with the environment reveal interesting properties of the system's evolution. Different approaches to estimate the bound on the speed of evolution of open quantum systems have been discussed in the literature. Quantum speed limit for open quantum systems in terms of quantum Fisher information and relative purity was derived in~\cite{taddei2013}, and~\cite{del2013}, respectively. In~\cite{deffner2013quantum}, geometric generalization
of both the Mandelstamm–Tamm and the Margolus–Levitin bounds was derived. In~\cite{wu2018, wu2020quantum} tighter bounds on speed limit time for mixed initial states were defined. In the present work, we consider the above-described approaches to bounds on the speed limit for open quantum systems in the context of quantum non-Markovianity. \newline

Different measures are made use of to quantify the impact and the memory effects of non-Markovian quantum systems. Quantum speed limit time is one among them. Quantum speed limit as a measure of non-Markovianity, and its connection with the hierarchy of quantum correlations of composite quantum system ~\cite{Paulson2014,paulson2017} are discussed in ~\cite{teittinen2019,paulson2021}. 
Even though quantum speed limit time seems to identify the information backflow due to the non-Markovianity, in~\cite{teittinen2019} it is shown that there exists no simple connection between non-Markovianity and speed limit time. This raises the question of whether there is a quantum concept of memory that is better indicated by the quantum speed limit. In this context, here we consider the quantification of memory as a deviation from `temporal self-similarity' \cite{shrikant2020}, which roughly refers to the idea of the form of the intermediate map being independent of the initial time. It is known to be equivalent to the dynamical semigroup and thus provides a concept of non-Markovianity weaker than non-divisibility and information backflow.  \newline

In this work, we consider the effect of non-Markovianity of single-qubit channels on the evolution of quantum states. We use both unital and non-unital maps and show that the quantum memory can speed up the quantum evolution. The quantum speed limit calculations based on relative purity, quantum Fisher information, and geometrical distance are made, and the dependence on quantum memory is established.\newline

The present work is organized as follows. Sec. II discusses the quantification of non-Markovianity as a deviation from the temporal self-similarity and quantum speed limit time based on relative purity. Sec. III discusses the effect of quantum non-Markovianity on speed limit time, and the cases of both dissipative and dephasing channels are considered.  A tighter bound on quantum speed limit time based on Bures angle is defined, and the effect of memory on it is given in Sec. IV. In Sec. V, we consider the speed limit time for mixed initial states, the concluding remarks in Sec. VI. In the literature, for  bounds on the evolution speed of quantum states, both quantum speed limit and quantum speed limit time are used interchangeably. In the main work, we present quantum speed limit time as a function of memory. We illustrate the quantum speed limit as a function of quantum Fisher information in the appendix (Sec.VII) for consistency and completeness.

\section{Preliminaries}
Here we briefly discuss the time-local master equation which would be used to generate the non-Markovian channels. We also lay down the recently developed measure of non-Markovianity based on temporal self-similarity.

\subsection{Time Local Master Equation}
The master equation local in time for a $d$-dimensional quantum system can be canonically written in the form,
\begin{align}
 \dot{\rho}=\mathcal{L} (\rho_{t})&=\frac{-i}{\hbar}[H(t),\rho_{t}]\nonumber \\
   &+\sum_{\mu=1}^{d^2-1}\gamma_{\mu}(t)\bigg[ L_{\mu}(t)\rho_{t} L_{\mu}^{\dag}(t) -\frac{1}{2}\{L_{\mu}^{\dag}(t)L_{\mu}(t),\rho_{t}\}\bigg]\,,
   \label{cmeq}
   \end{align}
where $\{L_{\mu}(t)\}$ forms an orthonormal basis set of traceless operators, i.e., $\textrm{tr}[L_{\mu}(t)]=0$,   $\textrm{tr}[L_{m}^{\dag}(t)L_{n}(t)]=\delta_{mn}$, and $H(t)$ is a Hermitian operator. Also, $\gamma_{\mu}(t)$ and $L_{\mu}(t)$ are the time dependant decoherence rates and decoherence operators, respectively. The decoherence rate $\gamma_{\mu}(t)$ is uniquely  (canonical decoherence rates, obtained by diagonalizing the Kossakowski matrix that appears in the general GKSL equation) defined and invariant under unitary transformations. The value  of $\gamma_{\mu}(t)$ determines the nature of the interaction of a system with its environment. If the decoherence rate is positive, the quantum channel is divisible, i.e., a quantum channel can be written as a concatenation of two non-unitary channels.  On the other hand, if $\gamma_{\mu}(t)$ is negative, then the evolution would be CP-indivisible. 
\newline
The memoryless master equation of Linblad form under Born-Markov and rotation wave approximations is, $\dot{\rho}=\frac{-i}{\hbar}[H,\rho]
   +\sum_{\mu=1}^{d^2-1}\gamma_{\mu}\bigg[ L_{\mu}\rho L_{\mu}^{\dag} -\frac{1}{2}\{L_{\mu}^{\dag}L_{\mu},\rho\}\bigg]
   $, $L_\mu$ and $\gamma_\mu$ are Linblad operators and rate constant, respectively.
\subsection{Measure of non-Markovianity}
In~\cite{shrikant2020} a measure of non-Markovianity as a deviation from temporal self-similarity was defined as,
\begin{equation}
    \zeta=\min_{\mathcal{L}^{*}}\frac{1}{T} \int_{0}^{T}\vert\vert \mathcal{L}(t)-\mathcal{L}^{*}\vert\vert dt,
    \label{NMMSR}
    \end{equation}
where, $\vert\vert A\vert\vert=\textrm{tr}\sqrt{A A^\dag}$ is the trace norm of the operator,  $\mathcal{L}(t)$ and $\mathcal{L}^{*}$ are the  generators of non-Markovian and Markovian evolutions, respectively. $\zeta=0$ iff the channel is a quantum dynamical semigroup (QDS) and is greater than zero for a deviation from QDS. \newline  

\subsection{Quantum Speed Limit}  
Quantum speed limit time defines a  bound  on the  minimum  time  required for a quantum system to evolve between two states. This is estimated in the case of open quantum systems using different distance measures for quantum states. Thus, for example, there is the bound analogous to the MT bound based on the relative purity~\cite {del2013} for open quantum systems. Another MT type bound, in which speed limit is derived in terms of variance of the generator, was obtained as a function of quantum Fisher information for non-unitary evolution~\cite{taddei2013}. MT and ML type bounds on speed limit time based on a geometrical distance between the initial and final states, which is a tighter bound, was developed in~\cite{deffner2013}. We study quantum speed limit time as a function of memory for pure as well as mixed initial states.\newline

\section{Impact of quantum memory on quantum speed limit time}

Here we examine the effect of memory on the speed of quantum evolution for various quantum processes. Dephasing \cite{banerjee2007} and dissipative processes are taken into consideration.  We consider divisible and indivisible non-Markovian quantum maps \cite{ghosal2021}, and connections between memory and quantum speed limit time are established. To this end, initially, the case of the CP divisible model is considered. We begin with the MT bound based on relative purity. A bound to the required time of evolution  analogous to the MT bound based on the relative purity~\cite {del2013} for open quantum system in which reference is made to the initial state and the dynamical map is,
    \begin{equation}
  \tau\geq\tau_{QSL}=\frac{\vert \cos\theta-1\vert \textrm{tr}\rho_{0}^{2}}{\textrm{tr}[(\mathcal{L}^{\dag}\rho_{0})^2]}\geq\frac{4\theta^2\textrm{tr}\rho_{0}^2}{\pi^2\sqrt{\textrm{tr}[(\mathcal{L}^{\dag}\rho_{0})^2]}},
        \label{spdlmt}
    \end{equation}
where $\theta=\cos^{-1}[\mathcal{P}(t)]$ with $\theta\in[0,\pi/2]$, $\mathcal{P}(t)=\textrm{tr}(\rho_{t}\rho_{0})/\textrm{tr}(\rho_{0}^2)$ is the relative purity of initial and final states. Here, $v=\sqrt{\textrm{tr}[(\mathcal{L}^{\dag}\rho_{0})^2]}$ gives an upper bound to the speed of the evolution.
The generalization of time-dependent $\mathcal{L}(t)$ is 
\begin{equation}
  \tau\geq  \tau_{QSL}=\frac{4\theta^2\textrm{tr}\rho_{0}^2}{\pi^2\overline{\sqrt{\textrm{tr}[(\mathcal{L}^{\dag}\rho_{0})^2]}}}.
    \label{MT_RP}
\end{equation}
Here $\overline{X}=\tau^{-1}\int_{0}^{\tau}X dt$.

Interestingly, in some recent works \cite{funo2019,arpan2021}, a thermodynamic interpretation was provided for the terms which arise in the QSL equation in a scenario where where time-local master equations govern quantum systems.

\subsection{ Dephasing quantum channels} 
The dynamics of a quantum system under a dephasing process in the interaction picture is given by, 
\begin{equation}
    \dot{\rho_{t}}=\gamma(t)(\sigma_{z}\rho_{t} \sigma_{z}-\rho_{t}),
    \label{depmas}
\end{equation}
where $\sigma_z$ is the Lindblad operator for dephasing process and $\gamma(t)$ is the rate of dephasing.
The initial state
\begin{equation}
\rho_{0}=\frac{1}{2}
    \begin{pmatrix}
    1+r_{z} & r_{x}-i r_{y}  \\
  r_{x}+i r_{y}  & 1-r_{z}
\end{pmatrix},
\label{intstate}
\end{equation}
where $r=(r_{x},r_{y},r_{z})$, $r \epsilon \mathcal{R}^3 $, and $||r||\leq1$
evolves to
\begin{equation}
\rho_{t}=\frac{1}{2}
    \begin{pmatrix}
    1+r_{z} & (r_{x}-i r_{y}) p_{t}~  \\
  (r_{x}+i r_{y})p_{t}  & 1-r_{z}
\end{pmatrix},
\label{dephfinal}
\end{equation}
where $p_t=e^{-2\Lambda_{t}}$, $\Lambda_{t}=\int_{0}^{t}\gamma(t)dt$, decoherence rate $\gamma(t)=-\frac{\dot{p_t}}{2p_t}$, and $p_t$ is the decoherence function.
In order to make use of the measure of non-Markovianity $\zeta$ (eq. \ref{NMMSR}), we note that for the dephasing  process, Eq. (\ref{depmas}),
    $\mathcal{L}-\mathcal{L^*}=(\gamma^*-\gamma)(\vert\phi^+\rangle\langle\phi^+\vert-\vert\phi^-\rangle\langle\phi^-\vert)$, with $\vert\phi^{\pm}\rangle$ being the Bell diagonal states.\newline
Quantum speed limit in this case is,
\begin{equation}
    \tau_{QSL}=\frac{4\sqrt{2}\cos^{-1}(\mathcal{P})^2\textrm{tr}\rho_{0}^2}{\pi^2/\tau{\int_{0}^{\tau}\vert\frac{\dot{p_t}}{p_t}\sqrt{r_{x}^2+r_{y}^2}\vert dt}}.
    \label{qslt_rp_phd}
\end{equation}
Here $\mathcal{P}(t)=(1 + p_t (r_x^2 + r_y^2) + r_z^2)/(1 + r_x^2 + r_y^2 + r_z^2)$ is the relative purity and we have $2 \textrm{tr}\rho_{0}^2=1+r_{x}^2+r_{y}^2+r_{z}^2$. Non-Markovianity and speed limit are calculated for different decoherence functions corresponding to CP-divisible and indivisible quantum channels. Their details are given below.
\subsubsection{CP-divisible phase damping channel; Ornstein–Uhlenbeck noise (OUN)}
OUN noisy channel even though Markovian  from the perspective of the CP-divisibility criteria, is non-Markovian ~\cite{yu2010,shrikant2020}.
The decoherence function of OUN is \cite{pradeep},
\begin{equation}
    p_t=  e^{\frac{-\mu}{2}\{t+\frac{1}{\Gamma}(e^{-\Gamma t}-1)\}},
    \label{OUNfn}
\end{equation}
where, $\Gamma^{-1}\approx\tau_{r}$ defines reservoir's finite correlation time and $\mu$ is the coupling strength related to qubit's relaxation time.
The decoherence rate is,
\begin{equation}
    \gamma(t) =
          \frac{\mu (1-e^{-\Gamma t})}{4}.
\end{equation}
This  channel is not CP-indivisible, $\gamma(t)$ is positive for all values of $t$. Nevertheless, it is non-Markovian, according to the measure Eq.~(\ref{NMMSR}), by virtue of its deviation from the QDS. The Markovian regime is achieved in the limit $\frac{1}{\Gamma}\rightarrow\infty$, with the corresponding decoherence function being $p^{*}(t)=e^{-\mu t/2}$. The quantum speed limit time, as a function of non-Markovianity, is calculated for the initial state $\frac{1}{\sqrt{2}}(\vert 0\rangle+\vert 1\rangle)$, and is depicted in Fig. \ref{qds_qslt_amd}.

\subsubsection{P-indivisible phase damping; Random Telegraph noise (RTN)}
RTN channel~\cite{mazzola2011}, is non-Markovian according to the information backflow and CP-divisibility criteria. The decoherence function in this case \cite{pradeep} has the form $p_t=e^{-\mu t}\Big[\cos\Bigg(\sqrt{[(\frac{2a}{\mu})^2-1]}\mu t\Bigg)+\frac{\sin\Bigg(\sqrt{[(\frac{2a}{\mu})^2-1]}\mu t\Bigg)}{\sqrt{(\frac{2a}{\mu})^2-1}}\Big]$.
The parameters $a$ and $\mu$ correspond to the strength of the system-environment coupling and the fluctuation rate of the RTN, respectively. There are two regimes of  system dynamics. For $\frac{a}{\mu}<0.5$, the channel corresponds to  the purely damping regime of Markovian dynamics, and damped oscillations for $\frac{a}{\mu}>0.5$ correspond to the non-Markovian evolution.
\begin{figure}[htbp]
    \centering
    \includegraphics[height=65mm,width=1\columnwidth]{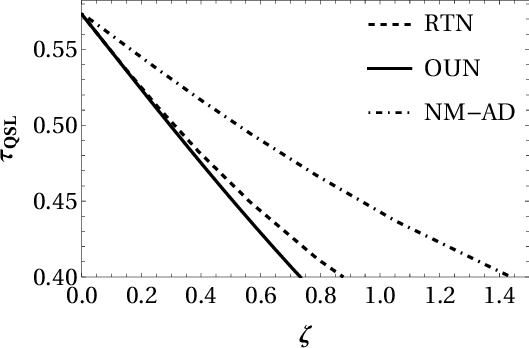}
    \caption{Quantum speed limit time $\tau_{QSL}$ is plotted as a function of measure of non-Markovianity $\zeta$ for the case OUN, RTN (Eq.\ref{qslt_rp_phd}) and NMAD (Eq.\ref{qslt_rp_amd}) channels. Quantum speed limit time $\tau_{QSL}$ is estimated for the initial states $\frac{1}{\sqrt{2}}[\vert0\rangle+\vert 1\rangle]$ for OUN and RTN channels, respectively and $\vert1\rangle\langle1\vert$ for NMAD. The channel parameter  $\Gamma=0.1\mu$, $\Gamma=\frac{1}{4}\mu$ and $\frac{a}{\mu}=1$ for NMAD, OUN and RTN, respectively for an actual driving time $\tau=1$.}
    \label{qds_qslt_amd}
\end{figure}

Figure \ref{qds_qslt_amd} depicts that quantum speed limit time decreases as  memory increases, indicated by increase in the value of $\zeta$, for an initial pure state, as indicated in the figure caption. 

\subsection{Amplitude damping channel}
After the two unital channels discussed above, we now consider a non-unital channel comprising of the  Jaynes-Cummings model for a two-level system resonantly coupled to a leaky single-mode cavity. The non-unitary generator of the reduced dynamics of the system is
\begin{equation}
\dot{\rho}=\gamma_{i}\Bigg(\sigma_{-}\rho_{t}\sigma_{+}-\frac{1}{2}\sigma_{+}\sigma_{-}\rho_{t}-\frac{1}{2}\rho_{t}\sigma_{+}\sigma_{-}\Bigg),
\end{equation}
where $\sigma_{\pm}=\frac{1}{2}(\sigma_{x}\mp i\sigma_{y})$ are the atomic raising and lowering operators. The emission is described by the Linblad operator $\sigma_{-}$, and $\gamma(t)$ is the rate of dissipation.
For the initial state, Eq.~(\ref{intstate}), the time evolved reduced density matrix becomes,
\begin{equation}
\rho_{t}=\frac{1}{2}
    \begin{pmatrix}
    2-(1-r_{z})\vert p_{t}\vert^2 & (r_{x}-i r_{y})p_{t} \\
  (r_{x}+i r_{y})p_{t}  & (1-r_{z})\vert p_{t}\vert^2
\end{pmatrix},
\label{adfinal}
\end{equation}
where $p_{t}=e^{-\Lambda_{t}/2}$, $\Lambda_{t}=\int_{0}^{t}\gamma(t)dt$. Also, $\gamma(t)=-\frac{\dot{2p_{t}}}{p_t}$, and $p_t=e^{-\Gamma t/2}\big( \cosh(dt/2)+\frac{\Gamma}{d}\sinh(dt/2)\big)$ with the time dependent decoherence rate,
\begin{equation}
    \gamma(t)=\frac{2\mu\Gamma \sinh(dt/2)}{d \cosh(dt/2)+\Gamma\sinh(dt/2)}.
\end{equation}
Here $d=\sqrt{\Gamma^2-2\mu\Gamma}$, $\Gamma$ is the spectral width of the reservoir, and $\mu$ is the coupling strength between the qubit and the cavity field.\newline
The measure of non-Markovianity $\zeta=\min_{\gamma^{*}}\frac{1}{\tau}\int_{0}^{\tau}\vert\gamma(t)-\gamma^{*}\vert(1+\sqrt{2})dt$. For this scenario, the quantum speed limit is calculated as,
\begin{equation}
\small
    \tau_{QSL}=\frac{4\sqrt{2}\cos^{-1}(\mathcal{P})^2\textrm{tr}\rho_{0}^2}{\pi^2/\tau\int_{0}^{\tau}\vert\frac{\dot{p_t}}{p_t}\sqrt{r_{x}^2+r_{y}^2+4(1+r_{z}^2)}\vert dt}.
    \label{qslt_rp_amd}
\end{equation}
The relative purity is 
    $\mathcal{P}(t)=(1-r_z + p_t(r_x^2 + 
    r_y^2 + p_t r_z (1 + r_z))/(1 + r_x^2 + r_y^2 + r_z^2)$. The behavior of quantum speed limit with memory, for the present case, is depicted in fig. \ref{qds_qslt_amd}.

\section{Bures angle and speed limit time}
MT and ML-type bounds on speed limit time  are estimated  by availing the geometrical approaches to quantify the closeness between the  initial and final states. Here, Bures angle is used to measure the distance between two quantum states. In~\cite{deffner2013quantum}, for the initial pure state $\rho_0=\vert\psi_0\rangle\langle\psi_0\vert$, a bound on the speed limit time based on Bures angle $\mathcal{B}(\rho_0,\rho_t)$ is,
\begin{equation}
    \tau_{QSL}=\max\Bigg\{\frac{1}{\Lambda^{\textrm{op}}_{\tau}},\frac{1}{\Lambda^{\textrm{tr}}_{\tau}},\frac{1}{\Lambda^{\textrm{hs}}_{\tau}}\Bigg\} \sin^2[\mathcal{B}],
    \label{B_spdlmt}
\end{equation}
where $\mathcal{B}(\rho_{0},\rho_{t})=\arccos\sqrt{\mathcal{F}(\rho_{0},\rho_{t})}$, and the Bures fidelity $\mathcal{F}(\rho_{0},\rho_{t})$ is defined as
\begin{equation}
    \mathcal{F}(\rho_{0},\rho_{t})=\bigg[\textrm{tr}[\sqrt{\sqrt{\rho_{0}}\rho_{t}\sqrt{\rho_{0}}}]\bigg]^{2}.
    \label{U_fidelity}
\end{equation}
Here, $\frac{1}{\Lambda^{\textrm{op}}_{\tau}}$,$\frac{1}{\Lambda^{\textrm{tr}}_{\tau}}$, and $\frac{1}{\Lambda^{\textrm{hs}}_{\tau}}$ are operator, Hilbert-Schmidt and trace norms, respectively, and
\begin{equation}
    \Lambda^{\textrm{op,tr,hs}}_{\tau}=\frac{1}{\tau}\int^{\tau}_{0} dt \vert\vert \mathcal{L}(\rho_t)\vert\vert_{\textrm{op,tr,hs}}.
\end{equation}
It is known that the operators satisfy the following inequality 
$\vert\vert A\vert\vert_{\textrm{op}}\leq\vert\vert A\vert\vert_{\textrm{hs}}\leq\vert\vert A\vert\vert_{\textrm{tr}}$. As a result, $1/\Lambda^{\textrm{op}}_{\tau}\geq1/\Lambda^{\textrm{hs}}_{\tau}\geq1/\Lambda^{\textrm{tr}}_{\tau}$, which shows that quantum speed limit time based on operator norm of the nonunitary generator provides the tighter bound on $\tau_{QSL}$. The computability of fidelity in Eq. (\ref{U_fidelity}) is analytically cumbersome. A simpler expression for an upper bound on fidelity is given  in~\cite{miszczak2008sub}, and it shows that for any density matrices $\rho_1$ and $\rho_2$, $\mathcal{F}(\rho_1,\rho_2)\leq \textrm{tr}\rho_1\rho_2+\sqrt{(1-\textrm{tr}\rho_1^2)(1-\textrm{tr}\rho_2^2)}$, with equality for single qubit. Making use of this super-fidelity, a modified bound on quantum speed limit time for both pure and mixed initial states is derived~\cite{wu2020quantum}. And speed limit time $\tau_{QSL}$ (Eq.~\ref{B_spdlmt}) can be estimated by modifying the denominator as,
\begin{equation}
    \Lambda^{\textrm{op,tr,hs}}_{\tau}=\frac{1}{\tau}\int^{\tau}_{0}dt\vert\vert \mathcal{L}(\rho_t)\vert\vert_{\textrm{op,tr,hs}}\Bigg(1+\sqrt{\frac{1-\textrm{tr}\rho_0^2}{1-\textrm{tr}\rho_t^2}}\Bigg).
\end{equation}

\begin{figure}[htbp]
    \centering
    \includegraphics[height=65mm,width=1\columnwidth]{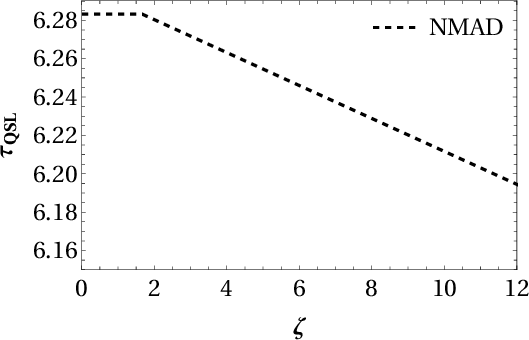}
    \caption{Quantum speed limit time  $\tau_{QSL}$ (Eq.\ref{amd_pure_bures}) is plotted as a function of the measure of non-Markovianity $\zeta$ for the NMAD channel.  $\tau_{QSL}$ is estimated for the initial state  $\vert1\rangle\langle1\vert$. The channel parameter $\Gamma=0.1\mu$ and actual driving time $\tau=2\pi$.}
    \label{tss_b_amd}
\end{figure}
For phase damping channel the Bures angle based quantum speed limit time is estimated as,
\begin{equation}
    \tau_{QSL}=\frac{1-p_t(r_x^2+r_y^2)-r_z^2-l_1 l_{2 t}}{\frac{1}{\tau}\int_0^\tau dt\vert\dot{p_t}\sqrt{r_x^2+r_y^2}(1+\frac{l_1}{l_{2t}})\vert},
\end{equation}
where $l_1=\sqrt{1-(r_x^2+r_y^2+r_z^2)}$, and $l_{2t}=\sqrt{1-p_t^2(r_x^2+r_y^2)-r_z^2)}$.
Similarly, for amplitude damping channels for the states in Eq. (\ref{intstate}) and (\ref{adfinal}),
\begin{equation}
    \tau_{QSL}=\frac{1+r_{z}-p_{t}(r_{x}^2+r_{y}^2+p_{t}r_{z}(1+r_{z}))-h_{1}h_{2t}}{\frac{1}{\tau}\int_{0}^{\tau}dt\vert\dot{p_{t}} \sqrt{r_{x}^2 + r_{y}^2 + 4 p_{t}^2 (1 + r_{z})^2}(1+\frac{h_1}{h_{2t}})\vert},
    \label{amd_pure_bures}
\end{equation}

where we have $h_{1}=\sqrt{1-(r_{x}^2+r_{y}^2+r_{z}^2)}$, and
$h_{2t}=\sqrt{p_{t}^2(2-r_{x}^2-r_{y}^2+2 r_{z}-p_{t}^2 (1+r_z)^2)}$. For  pure states  $\sum r_{i}^2=1$. Figure~\ref{tss_b_amd} depicts $\tau_{QSL}$ as a function of memory for initial excited state $(0,1)^T$ in a dissipative process. It is evident that speed limit time decreases as memory increases.


\section{speed limit time for mixed initial states}
\begin{figure}
    \centering
    \includegraphics[height=65mm,width=1\columnwidth]{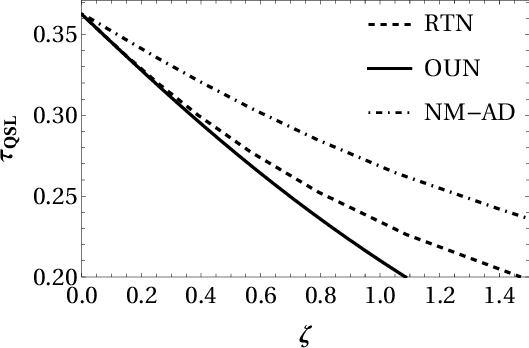}
    \caption{Quantum speed limit time  $\tau_{QSL}$ is plotted as a function of  $\zeta$ for OUN, RTN (Eq.~\ref{qslt_rp_phd}), and NMAD (Eq.~\ref{qslt_rp_amd}) channels, for mixed initial states, $r_{x}=\frac{1}{2},r_{y}=r_{z}=0$ (OUN and RTN) and $r_{x}=r_{y}=0,r_{z}=-\frac{1}{2}$ (NMAD).The channel parameter  $\Gamma=0.1\mu$, $\Gamma=\frac{1}{4}\mu$ and $\frac{a}{\mu}=1$ for NMAD, OUN and RTN, respectively for an actual driving time $\tau=1$.}
    \label{tss_mixed}
\end{figure}

We finally discuss the case for mixed initial states.  MT type bound based on relative purity (Eq.~\ref{MT_RP}) for initial mixed states, is shown in~fig.~\ref{tss_mixed}. As  seen for pure initial states (fig.~\ref{qds_qslt_amd}), quantum speed limit time decreases as memory increases for mixed initial states as well. 

The quantum speed limit's calculations based on Bures angle for mixed initial states are quite cumbersome. In ~\cite{campaioli2018} a combined  bound on the quantum speed limit time was obtained for almost all  mixed initial states by using a function of relative purity for unitary driven systems. Using the same function of relative purity, an easy to calculate  bound for initial mixed state for the open quantum system was subsequently derived in~\cite{wu2018},
\begin{equation}
    \tau_{QSL}=\max\Bigg\{\frac{1}{\Lambda^{\textrm{op}}_{\tau}},\frac{1}{\Lambda^{\textrm{tr}}_{\tau}},\frac{1}{\Lambda^{\textrm{hs}}_{\tau}}\Bigg\} \sin^2[\mathcal{\phi}]\textrm{tr}\rho_{0}^2.
    \label{m_spdlmt}
\end{equation}
Here, $\cos\mathcal{\phi}$ is defined as the square root of relative purity $(\cos\mathcal{\phi}=\sqrt{\mathcal{P}(t)})$. Quantum speed limit time bound for dephasing channels  is,
\begin{equation}
    \tau_{QSL}=\frac{(1-p_t)(r_x^2+r_y^2)}{\frac{1}{\tau}\int_0^\tau dt\vert \dot{p_t} \sqrt{r_x^2+r_y^2}\vert}.
\end{equation}
For a  mixed initial state $r_x=\frac{1}{2},r_y=r_z=0$, quantum speed limit time for OUN and RTN quantum channel is a constant for the range of memory considered.\\
Expression of quantum speed limit time for states Eq. ~(\ref{intstate}) and ~(\ref{adfinal}) for NMAD channel can be shown to be,
\begin{equation}
    \tau_{QSL}=\frac{(1 -p_{t}) [r_{x}^2 + r_{y}^2 +r_{z} (1 +p_{t}) (1 + r_{z})]}{\frac{1}{\tau}\int_{0}^{\tau}\vert\dot{p_{t}} \sqrt{r_{x}^2 + r_{y}^2 + 4 p_{t}^2 (1 + r_{z})^2}\vert dt}.
    \label{qslt_srp_amd}
\end{equation}

\begin{figure}
    \centering
    \includegraphics[height=65mm,width=1\columnwidth]{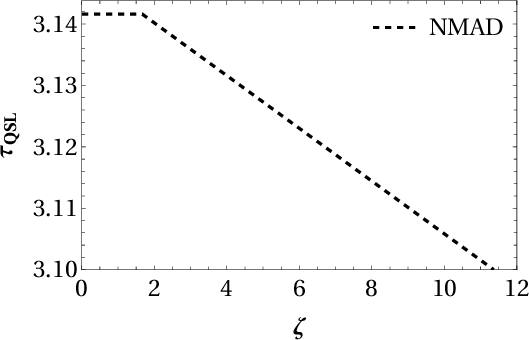}
    \caption{Quantum speed limit time  $\tau_{QSL}$, (Eq.~\ref{qslt_srp_amd}), is plotted as a function of $\zeta$ for the NMAD channel for mixed initial state ($r_{x}=r_{y}=0,r_{z}=-\frac{1}{2}$). The channel parameter  $\Gamma=0.1\mu$ and actual driving time $\tau=2\pi$.}
    \label{tss_amd_mixed}
\end{figure}

Quantum speed limit time is calculated with the initial mixed state $(r_{z}=-\frac{1}{2},r_{x}=r_{y}=0)$ for non-Markovian amplitude damping channel, and is depicted as a function of $\zeta$ in fig.~\ref{tss_amd_mixed}. As we have noticed in the previous cases, here also speed limit time decreases as memory increases.
\section{Conclusions}
We investigated the speed of evolution of open quantum systems with pure and mixed initial states. We considered CP-divisible and indivisible non-Markovian quantum channels, and showed the behavior of quantum speed limit as a function of quantum memory, quantified as the deviation from temporal self-similarity of quantum dynamical maps, which provides a weaker concept of non-Markovianity than CP-indivisibility.  We estimated both Mandelstamm-Tamm (MT) and Margolous-Levitin (ML) types bound based on various distance measures between the quantum states viz; relative purity, Bures angle and quantum Fisher information. In the case of  CP-divisible (OUN), CP-indivisible (RTN) dephasing, and dissipative (NM-AD) channels, for the initial states considered, quantum speed limit time decreases as non-Markovianity increases. Hence, quantum memory can enhance the speed of evolution between the quantum states. This is true for both pure and mixed initial states considered, and this need not be true for all initial states,  for example  see ~\cite{paulson2022quantum}. These outcomes highlight the beneficial impact of memory on the dynamics of the system of interest, exemplified here by the quantum speed limit. Our results for the OUN channel demonstrated the enhancement of the speed of quantum evolution in the presence of quantum memory for a wider class of channels than indicated by the CP-indivisibility criterion.
\label{sec4}
\section*{Acknowledgement}
SB and RS acknowledge the support from the Interdisciplinary Cyber-Physical Systems (ICPS) programme of the Department of Science and Technology (DST), India,
Grant No.: DST/ICPS/QuST/Theme-1/2019/6 and DST/ICPS/QuST/Theme-1/2019/14, respectively. RS also acknowledges the support of DST, India, Grant No. MTR/2019/001516.

\section*{Appendix}
\setcounter{equation}{0}
\appendix*

An MT type bound, in which speed limit is derived in terms of variance of the generator, was obtained as a function of quantum Fisher information for non-unitary evolution~\cite{taddei2013}.
A bound on $\mathcal{B}(\rho_{0},\rho_{\tau})$ can be obtained in terms of the integral of the quantum Fisher information $F_{Q}(t)$ along the evolution path.
The Bures fidelity $\mathcal{F}$ (Eq.~\ref{U_fidelity}) is connected to the quantum Fisher information $F_{Q}(t)$ ~\cite{taddei2013},
\begin{equation}
    \mathcal{F}(t,t+dt)=1-(dt)^2 F_{Q}(t)/4+\mathcal{O}(dt)^3.
\end{equation}
Quantum Fisher information is defined by $F_{Q}(t)=\textrm{tr}[\rho(t)L^2(t)]$. Here, the Hermitian operator $\hat{L}(t)$ is known as the symmetric logarithimic derivative (SLD) operator. It is defined as $d\hat{\rho}(t)/dt=(\hat{\rho}(t)\hat{L}(t)+\hat{L}(t)\hat{\rho(t)})/2$.
\begin{figure}[!htbp]
    \centering
    \includegraphics[height=65mm,width=1\columnwidth]{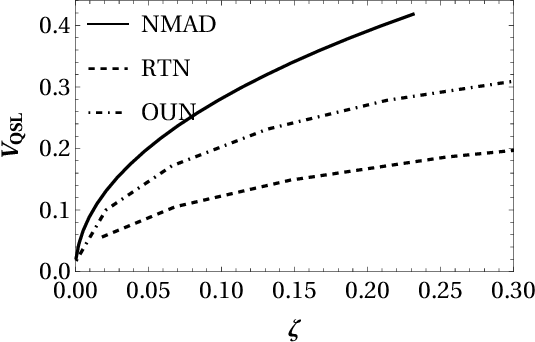}
    \caption{Quantum speed limit  $V_{QSL}$ is plotted as a function of the measure of non-Markovianity $\zeta$ for OUN, RTN and NMAD channels. Quantum speed limit $V_{QSL}$ is estimated for the initial states $\frac{1}{\sqrt{2}}[\vert0\rangle+\vert 1\rangle]$ for OUN and RTN channels, and $\vert1\rangle\langle1\vert$ for NMAD.The channel parameter  $\Gamma=0.1\mu$, $\Gamma=\frac{1}{4}\mu$ and $\frac{a}{\mu}=1$ for NMAD, OUN and RTN, respectively for an actual driving time $\tau=1$.}
    \label{tss_vql_amd}
\end{figure}
The instantaneous speed of evolution between two time intervals is proportional to  the square root of quantum Fisher information. The upper bound on Bures angle is,
\begin{equation}
    \mathcal{B}(\rho_{0},\rho_{\tau})=\arccos(\sqrt{\mathcal{F}(\rho_{0},\rho_{\tau})})\leq\frac{1}{2}\int_{0}^{\tau}\sqrt{F_{Q}(t)} dt,
\end{equation}
and the quantum speed limit can be identified as,
\begin{equation}
    V_{QSL}=\frac{1}{2}\sqrt{F_{Q}(t)}.
\end{equation}
This bound is attained only if the evolution occurs on a geodesic, a condition for the MT bound for unitary evolution under a time-independent Hamiltonian. \newline
We estimate the quantum speed limit $V_{QSL}$ in terms of quantum Fisher information for CP-divisible and CP-indivisible maps, for various initial pure states; $\frac{1}{\sqrt{2}}(\vert 0\rangle+\vert 1\rangle)$ for OUN and RTN channels, and $\vert 1\rangle\langle1\vert$ for the non-Markovian amplitude damping channel. In fig.~\ref{tss_vql_amd}, the quantum speed limit is seen to increase as non-Markovianity increases for both CP-divisible and indivisible channels. Thus, the evolution speed of quantum states is seen to increase with the strength of non-Markovianity.
\section*{Data Availability}
All data generated or analysed during this study are included in this published article.

 
%

\end{document}